\documentclass[10pt]{article}
\usepackage{latexsym}
\usepackage[dvips]{color}
\usepackage{float}
\usepackage{afterpage} 
\usepackage [pdftex]{graphicx}
\usepackage{epsfig}
\usepackage{amsmath}
\usepackage{amsfonts}
\usepackage{amssymb}
\usepackage{setspace}
\usepackage{subfigure}

\title{John Cage's Number Pieces as Stochastic Processes: a Large-Scale Analysis}
\author{
        Alexandre Popoff \\
                al.popoff@free.fr\\
        France}

\begin{document}

\maketitle

\section{Introduction}

	Starting from 1987 to 1992, the composer John Cage began writing a series of scores named the "Number Pieces". The Number Pieces are easily identifiable through their titles, which refer to the number of performers involved, and the rank of the piece among those with the same number of performers. For example \textit {Four} is the first Number Piece written for four performers, whereas \textit{Four$^3$} is the third one. From 1987 till Cage's death, forty-seven such pieces were written.
	
	In all the Number Pieces except \textit{One$^3$} and \textit{Two}, John Cage used a particular time-structure for determining the temporal location of sounds which was named "time-bracket". These time-brackets already appeared in earlier works such as \textit{Thirty Pieces for Five Orchestras} and \textit{Music for...}. However, in the Number Pieces, Cage simplified the contents of the time-brackets, most of them containing only a single tone or sound, especially in his late works.

\begin{figure}
\centering
\includegraphics[scale=0.7]{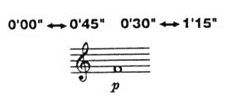}
\caption{A typical time-bracket from Cage's Number Piece \textit{Five}}
\label{fig:timebracketexample}
\end{figure}

	A time-bracket is basically made of three parts : a fragment of one or many staves, lying under two time intervals, one on the left and one on the right. A typical time-bracket can be seen on Figure \ref{fig:timebracketexample}. The time intervals consist of two real-time values separated by a two-way arrow. The staves contain one or more sound events without any duration indications. The time-bracket is performed as follow : the performer decides to start playing the written sounds anywhen inside the first time interval on the left, and chooses to end them anywhen inside the second one. These parameters are thus left free to the performer, provided he respects the time-bracket structure. In the example of Figure \ref{fig:timebracketexample}, the performer can start playing the note F whenever between 0 and 45 seconds, and can choose to end it whenever between 30 and 75 seconds (assuming of course that the note has started before). Note that there exists an overlap, which we will call the \textit{internal overlap}, between the starting time interval and the ending time interval, which is 15 seconds long in this specific case.
	
	In a few cases, a time-bracket may also be fixed. In that case, time intervals are replaced with single indications of time, for example 2'15'' and 2'45'', meaning that sounds should always begin and end at the indicated times.
	
	Successive time-brackets occurs in a Number Piece score with possible overlap between each other, which we will call \textit{external overlaps}, meaning that the ending time interval of one time-bracket may overlap the beginning interval of the next one.
	
	In the case of Number Pieces written for multiple performers, the superposition of different voices each playing time-brackets according to their choice creates a polyphonic landscape in constant evolution. Previous authors (\cite{weisser1}, \cite{weisser2}, \cite{haskins}) have shown how the Number Pieces were written as a consequence of Cage's new insights about harmony. In particular, Haskins (\cite{haskins}) has commented in a detailed dissertation on Cage's views about harmony throughout his career. Cage had been critical of traditional Western harmony, and even of twelve-tone methods, as they were based on rules which prevented the appearance of certain combination of pitches. A majority of his work arose as expressions of a "liberated" harmony where any sound or chord, or transition between them, could happen, mainly through chance operations. It is notable that, towards the end of his career, Cage seems to have adopted a conception of harmony which simply consists in the principle of sounds sounding together at the same time. To quote: \textit{"harmony means that there are several sounds...being noticed at the same time, hmm ? It's quite impossible not to have harmony, hmm ?"} (\cite{cage}).
	
	In his dissertation, Haskins points out (\cite{haskins}, p. 196) that the analysis of the Number Pieces is complex \textit{"...because the brackets offer a flexibility that creates many possibilities"}. He later adresses the same problem when analyzing \textit{Five$^2$}: \textit{"Coping with the myriad possibilities of pitch combinations - partially ordered subsets - within each time-bracket of Five$^2$ remains an important issue"} (\cite{haskins}, p. 207), and cites the work of Weisser on \textit{Four}. In \cite{weisser1} and \cite{weisser2}, Weisser enumerated the possible pitch-class sets in \textit{Four}, classifying them in \textit{"certain triads/seventh chords"}, \textit{"possible triads/seventh chords"}, \textit{"thwarted triads/seventh chords"} and \textit{"triadic segments"}. By doing so, Weisser is able to identify some of the possible pitch-class sets which can occur during a time-bracket. However, his analysis presents some drawbacks. The first one is that Weisser concentrates on triads and seventh chords and neglects the other possible pitch-class sets. As will be seen below, a performance of \textit{Four} opens the possibility of hearing 49 different pitch-class sets (including silence, single sounds and dyads). The second drawback is that Weisser's analysis poorly takes into account the inherently random temporal structure of the time-bracket, in which sounds from different parts may begin and end at different times. For example, he classifies (\cite{weisser2}, p. 202, Example 10a) a seventh chord in the last time-bracket of \textit{Four}, section C, as \textit{"virtually certain"} chord. However, if one of the player stops playing before the others have entered, as is possible given the rules of time-brackets interpretation, then this chord will not be heard. Haskins faced the same difficulties and, in the case of \textit{Five$^2$}, turned to the analysis of one particular performance taken from a recording.
	
	In previous works (\cite{popoff1}, \cite{popoff2}), we have advocated a statistical approach to the analysis of the time-bracket structure, focusing on a single time-bracket containing a single pitch. This approach allows to deal with the entire possibilities offered in terms of starting and ending times (and thus durations and temporal location in the time-bracket). The purpose of this paper is therefore to extend this approach to the analysis of an entire Number Piece, by considering all time-brackets and all parts. The determination of an entire part allows the determination of its sonic content over time. Having this information for each part allows the determination of the chords occuring during a performance. We use here the methods of (musical) set theory in which chords are identified by their corresponding pitch-class set. By averaging over a large number of realizations (which is achieved through a computer program running the determination of the parts repeatedly) we can access the probability distributions of each pitch-class set over time, thus turning the Number Pieces into stochastic processes. By doing so, we solve the problem posed by Haskins and Weisser of coping with all the possibilities offered by the Number Pieces. We have chosen to focus on two Number Pieces, namely \textit{Four} and \textit{Five}, as these are short pieces with a reduced number of players which therefore allows for a convenient computer implementation. This will also allow us to compare the results about \textit{Four} with Weisser's analysis. Section 2 of this paper describes the methodology used for the analysis, while section 3 and 4 present the analysis of \textit{Five} and \textit{Four} respectively.

\section{Methodology}

\subsection{General overview}

A \textit{part} in a Number Piece is the set of all time-brackets and their pitch content associated with a player's score. Given the score of a Number Piece, i.e the description of all time-brackets and their pitch content, we call \textit{realization of a Number Piece} the knowledge of starting and ending times for all pitches, after their selection from the time-brackets. .

In order to study a Number Piece as a whole from a statistical point of view, a computer program wa written in order to generate a large number $N$ (typically $10^4-10^5$) realizations of the Number Piece, and derive probabilities for the possible pitch-class sets. The programs for the analysis of \textit{Five} and \textit{Four} are written in ANSI C, mainly for speed issues. We give here the general overview of the program, while the specifics pertaining to particular choices in the algorithm will be given in the following subsections.

For each realization, the programs successively and independently generate the parts corresponding to each player. Whether parts in actual performances of the Number Pieces are indeed independent, or should be chosen so, is worth questioning. In commenting the specific example of \textit{Seven$^2$}, Weisser (\cite{weisser1}) underlines the fact that performers should work cooperatively to fullfill Cage's instructions. Nevertheless, we have chosen to select each part independently as it is practical (such a choice is the easiest to implement algorithmically), and for lack of a proper model of human behavior, which would be difficult to describe in such a case.

Once parts have been selected, the pitch-class content is known for each time $t$ in each part. By using Starr's algorithm (see below), we can thus derive the corresponding pitch-class set at each time $t$.

By averaging over $N=10^4-10^5$ realizations, it is possible to derive the probabilities $Pr(PCS_t =i)$ of obtaining the pitch-class set $i$ at time $t$. In other words, we obtain a collection of random variables $PCS_t$ indexed over time, with values in the possible pitch-class sets. In the framework of a statistical analysis, we thus see that there is a stochastic process naturally associated with the Number Piece being studied.  In the rest of this paper, the probabilities $Pr(PCS_t =i)$ will be presented under the form of a heat map with respect to time and possible pitch-class sets. Moreover, since we have access for each realization to the entire pitch-class set content over time, we can derive conditional probabilities of the form $Pr(PCS_{t+\tau} = j | PCS_{t} = i)$, $\tau > 0$. These probabilities are calculated using

$$Pr(PCS_{t+\tau} = j | PCS_{t} = i) = \frac{Pr(PCS_{t+\tau} = j \cap PCS_{t} = i)}{Pr(PCS_{t} = i)}$$

assuming that $Pr(PCS_{t+\tau} = j | PCS_{t} = i) = 0$ if $Pr(PCS_{t} = i)$ for continuity. These conditional probabilities are useful to determine the possible evolutions of pitch-class sets during a realization, as they express the probability of having pitch-class set $j$ at $\tau$ time in the future, knowing that we have pitch-class set $i$ in the present. 

\subsection{Time-bracket selection}

We have described in \cite{popoff1} the analysis of a single time-bracket containing a unique pitch, using  a temporal selection procedure based on uniform distributions over the starting and ending time intervals. In \cite{popoff2} we have studied the effect of using different distributions on the characteristics of the sounds thus produced. We will use in this analysis a similar strategy, as described below.

In all cases, we have discretized time, using a smallest unit of $0.1$ seconds. Though the indications of Cage for the time-brackets refer to a continuous time process, we feel that discretization is not an issue for analysis. Indeed, performers would probably use a stopwatch or a clockwatch in real performances, which usually have a maximum resolution of $0.01$ seconds. The devices used for time measurement thus already discretize time, and it is very unlikely that normal performers are able to choose starting and ending time with a $0.01$s precision. Hence a resolution of $0.1$ seconds seems a reasonable choice for analysis. From now on, all time indications will be given in tenth of seconds.

For a given part, time-brackets are processed successively in the order given on the score, and for each time-bracket starting times are determined before ending times, in order to mimick the situation of actual performers. The selection of starting and ending times for a given time-bracket follows a similar procedure to the one exposed in \cite{popoff1, popoff2}, with the addition that we have to account for possible external overlaps between time-brackets. For example, assume that, for a given part, a time-bracket has an ending time interval of $[300,750]$ and that the following time-bracket has a starting time interval of $[600,1050]$. If the pitch of the first time bracket finishes at $t=720$, then the starting time of the pitch of the second time-bracket cannot be selected inside $[600,1050]$ but should be selected in $[720,1050]$ instead.

For a given time interval, the selection procedure then boils down to the random selection of a point according to a given distribution $Pr(T=t)$ and depending on a parameter $t_{Prec}$. This parameter is either 

\begin{itemize}
\item{The previously chosen starting time in the same time-bracket, if the time interval is an ending time interval, or}
\item{the chosen ending time in the previous time-bracket, if the time interval is a starting time interval.}
\end{itemize}

\begin{figure}
\centering
\includegraphics[scale=0.3]{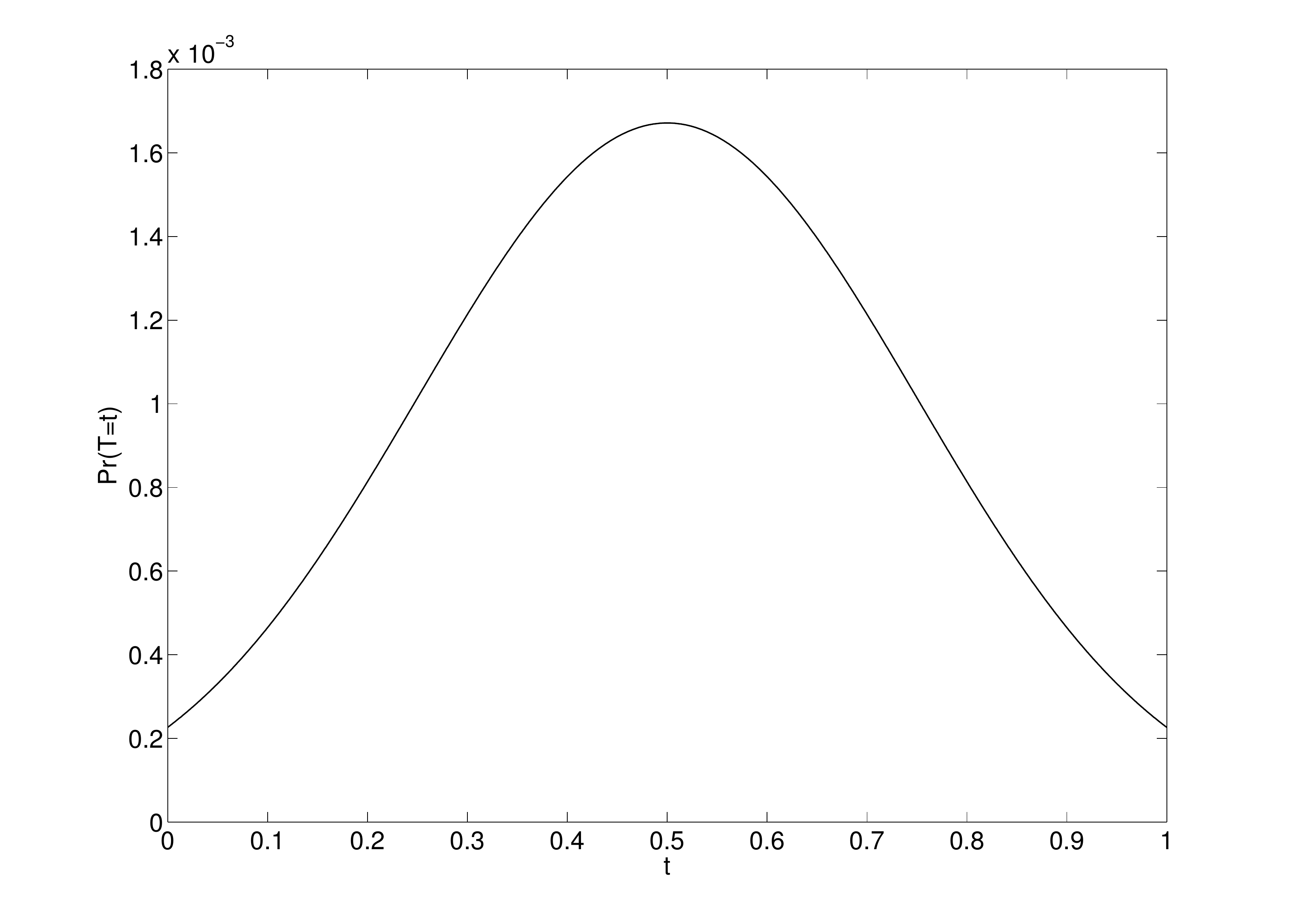}
\caption{The gaussian distribution used for selecting starting and ending times in their corresponding intervals. The distribution is presented here on a unit time interval with resolution 0.001 seconds. It is scaled accordingly depending on the time interval considered.}
\label{fig:gaussiandistribution}
\end{figure}

Thus, if the considered time interval is $[t_s,t_e]$ and $t_{Prec} < t_s$, the random selection will be performed on the interval $[t_s,t_e]$, otherwise we replace it with $[t_{Prec},t_e]$.

For a time interval $[t_s,t_e]$, we assume here a gaussian probability distribution of the form

$$Pr(T=t) = A.e^{-\dfrac{(t-c)^2}{2\sigma^2}}, c=\dfrac{t_s+t_e}{2}, \sigma=\dfrac{t_e-t_s}{4}$$

where A is a constant chosen so that the distribution normalizes to unity. A plot of this distribution for a discretized unit time interval (with resolution 0.001 seconds) is presented on Figure \ref{fig:gaussiandistribution}. It is scaled accordingly depending on the time interval considered. This distribution differs from the uniform distribution considered in \cite{popoff1} as it gives less prevalence to sound events occuring at the very beginning or end of their time interval. However, we have verified that the results of the analysis of the Number Pieces do not depend on a large scale on the distribution chosen for the realization of the time-brackets.

For fixed time-brackets, the selection procedure is simplified as their starting time and ending time are automatically determined.

\begin{figure}
\centering
\includegraphics[scale=0.7]{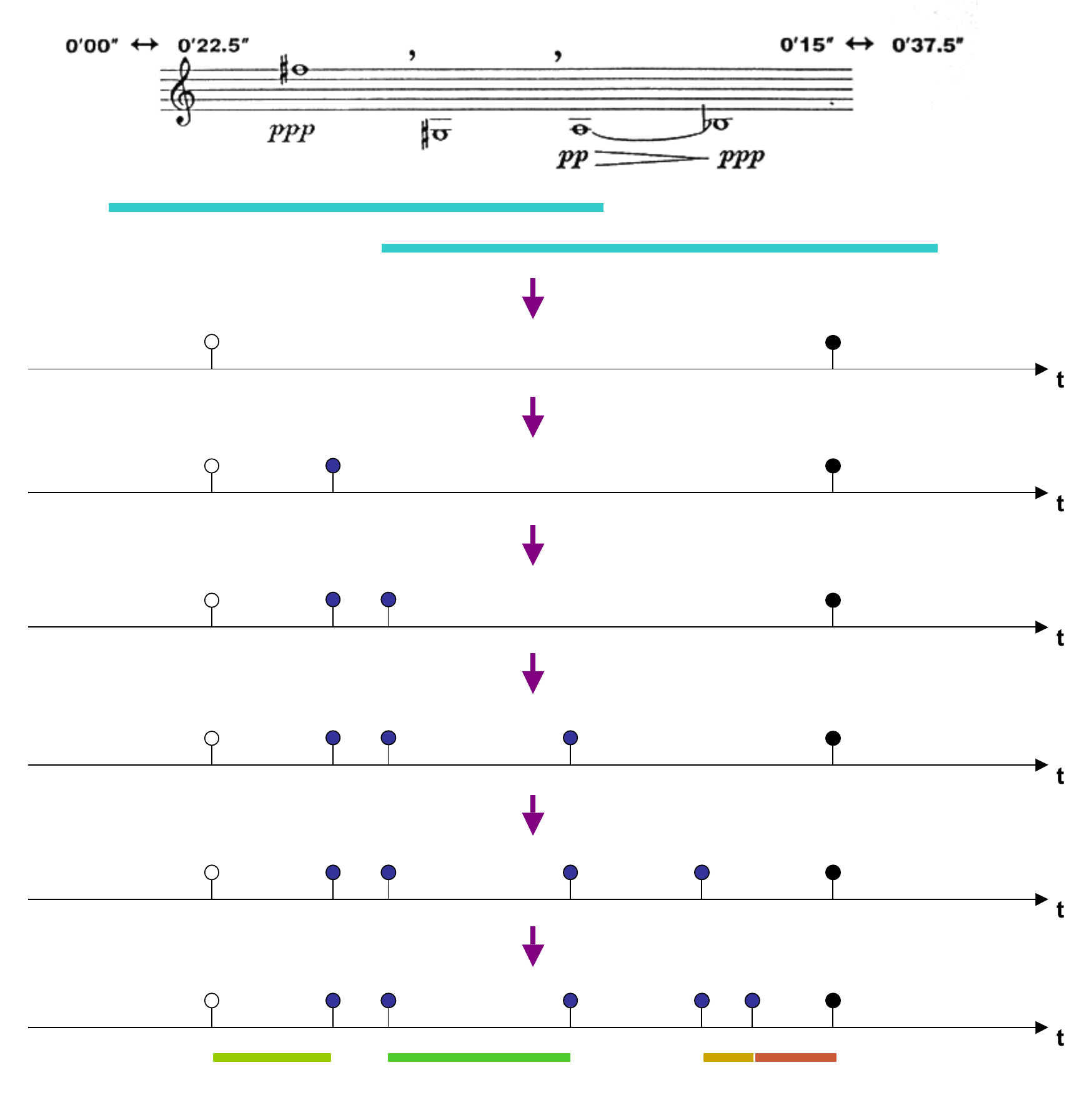}
\caption{The selection procedure for a complex time-bracket (top). The outer limits of the content are selected identically to a single pitch time-bracket. Additional time marks are then selected successively inside the obtained limits in order to determine the temporal location of each individual pitch (represented at the bottom of the figure by color bars).}
\label{fig:complexTBs}
\end{figure}

Some time-brackets may contain multiple pitches, as exemplified by the time-bracket on top of Figure \ref{fig:complexTBs}, wherein pitches F\#, G\# and A are separated by what appears to be pause indications, while pitches A and A\# are linked by a slur. In the absence of indications by Cage about the treatment of pauses, we have adopted the following procedure for the realization of complex time-brackets

\begin{enumerate}
\item{The outer limits $t_s$ and $t_e$ of the sonic content are selected in an identical way as they are for a single-pitch time-bracket. Additional time marks are then selected successively to determine the location of each pitch.}
\item{The presence of a pause indication necessitates the determination of two time marks. For example, to determine the temporal location of pitches F\# and G\# in the time-bracket of Figure \ref{fig:complexTBs}, a first time mark $t_1$ is selected in the interval $[t_s,t_e]$, following the distribution of Figure \ref{fig:gaussiandistribution}, then a second time mark $t_2$ is selected in the remaining interval $[t_1,t_e]$. A third selection in the time interval $[t_2,t_e]$ marks the end of G\#.}
\item{The presence of a slur indication necessitates the determination of only one time mark, as pitches are then heard without pause.}
\end{enumerate}

\begin{table}
{
\scriptsize
\begin{center}
\begin{tabular}{ | c | c | }
\hline
Extended Forte number & Pitch-class set \\ [1ex] 

\hline              

0-1 & $\emptyset$ \\
1-1 & $[0]$ \\
2-1 & $[0,1]$ \\
2-2 & $[0,2]$ \\
2-3 & $[0,3]$ \\
2-4 & $[0,4]$ \\
2-5 & $[0,5]$ \\
2-6 & $[0,6]$ \\
3-1 & [0,1,2] \\
3-2 & $[0,1,3]$ \\
3-3 & $[0,1,4]$ \\
3-4 & $[0,1,5]$ \\
3-5& $[0,1,6]$ \\
3-6& $[0,2,4]$ \\
3-7& $[0,2,5]$ \\
3-8& $[0,2,6]$ \\
3-9& $[0,2,7]$ \\
3-10& $[0,3,6]$ \\
3-11& $[0,3,7]$ \\
3-12& $[0,4,8]$ \\
4-1& $[0,1,2,3]$ \\
4-2& $[0,1,2,4]$ \\
4-3& $[0,1,3,4]$ \\
4-4& $[0,1,2,5]$ \\
4-5& $[0,1,2,6]$ \\
4-6& $[0,1,2,7]$ \\
4-7& $[0,1,4,5]$ \\
4-8& $[0,1,5,6]$ \\
4-9& $[0,1,6,7]$ \\
4-10& $[0,2,3,5]$ \\
4-11& $[0,1,3,5]$ \\
4-12& $[0,2,3,6]$ \\
4-13& $[0,1,3,6]$ \\
4-14& $[0,2,3,7]$ \\
4-z15 & $[0,1,4,6]$ \\
4-16& $[0,1,5,7]$ \\
4-17& $[0,3,4,7]$ \\
4-18& $[0,1,4,7]$ \\
4-19& $[0,1,4,8]$ \\
4-20	& $[0,1,5,8]$ \\
4-21	& $[0,2,4,6]$ \\
4-22	& $[0,2,4,7]$ \\
4-23	& $[0,2,5,7]$ \\
4-24	& $[0,2,4,8]$ \\
4-25	& $[0,2,6,8]$ \\
4-26	& $[0,3,5,8]$ \\
4-27	& $[0,2,5,8]$ \\
4-28	& $[0,3,6,9]$ \\
4-z29 & $[0,1,3,7]$ \\

\hline
\end{tabular}
\quad
\begin{tabular}{ | c | c | }
\hline
Extended Forte number & Pitch-class set \\ [1ex] 

\hline              
5-1& $[0,1,2,3,4]$ \\
5-2& $[0,1,2,3,5]$ \\
5-3& $[0,1,2,4,5]$ \\
5-4& $[0,1,2,3,6]$ \\
5-5& $[0,1,2,3,7]$ \\
5-6& $[0,1,2,5,6]$ \\	
5-7& $[0,1,2,6,7]$ \\
5-8& $[0,2,3,4,6]$ \\
5-9& $[0,1,2,4,6]$ \\
5-10& $[0,1,3,4,6]$ \\
5-11& $[0,2,3,4,7]$ \\
5-z12& $[0,1,3,5,6]$ \\
5-13& $[0,1,2,4,8]$ \\
5-14& $[0,1,2,5,7]$ \\
5-15& $[0,1,2,6,8]$ \\
5-16& $[0,1,3,4,7]$ \\
5-z17& $[0,1,3,4,8]$ \\
5-z18& $[0,1,4,5,7]$ \\
5-19& $[0,1,3,6,7]$ \\
5-20	& $[0,1,5,6,8]$ \\
5-21	& $[0,1,4,5,8]$ \\
5-22	& $[0,1,4,7,8]$ \\
5-23	& $[0,2,3,5,7]$ \\
5-24	& $[0,1,3,5,7]$ \\
5-25	& $[0,2,3,5,8]$ \\
5-26	& $[0,2,4,5,8]$ \\
5-27	& $[0,1,3,5,8]$ \\
5-28	& $[0,2,3,6,8]$ \\
5-29	& $[0,1,3,6,8]$ \\
5-30	& $[0,1,4,6,8]$ \\
5-31	& $[0,1,3,6,9]$ \\
5-32	& $[0,1,4,6,9]$ \\
5-33	& $[0,2,4,6,8]$ \\
5-34	& $[0,2,4,6,9]$ \\
5-35	& $[0,2,4,7,9]$ \\
5-z36& $[0,1,2,4,7]$ \\
5-z37& $[0,3,4,5,8]$ \\
5-z38& $[0,1,2,5,8]$ \\	

\hline
\end{tabular}

\end{center}
}
\caption{List of pitch-class sets up to pentachords with their corresponding extended Forte number}
\label{tab:pcsets}
\end{table}

This particular selection procedure may be questioned. For example, one could propose that all internal time marks be selected at once instead of successively, sorting them eventually to determine the temporal locations of pitches. We feel however that a successive procedure is more representative of performance behavior, as the musicians would focus only on one pitch at a time. We would like to note, however, that we have performed simulations with simultaneous sampling, instead of successive: the results show that, while the distributions of the random variables $PCS_t$ may be affected on short time-scales, the large scale behavior remains essentially the same. One may also object to the unequal treatment of the internal time marks as compared to the outer temporal limits. We have used this procedure to ensure that the determined time marks would respect the time-bracket structure. In any case, we are aware that other procedures exist, which could be applied for analysis. The remaining question of whether they accurately reflect human behavior is difficult, and has been adressed in \cite{popoff2}, noting in particular that human behavior may be too complex to model easily.

\subsection{Pitch-class set determination}

Given the knowledge of pitch-classes in each part at each time $t$, we determine the corresponding pitch-class set using Daniel Starr's algorithm, which has been described in \cite{starr1} and \cite{starr2}.
The pitch-class set are given using Forte's notation (\cite{forte}), which we extend to take into account silence (notated by 0-1), single sounds (1-1) and dyads (2-1 to 2-6). These Forte numbers
and their corresponding pitch-class sets can be found in Table \ref{tab:pcsets}. Note that the notation of pitch-class sets only differs from Forte's original notation for pitch-class set 5-20, which is $[0,1,3,7,8]$ in Forte's list. Note that we do not differentiate between a pitch-class set and its inverted form. Hence pitch-class set may designate either a major triad or a minor one.

\subsection{Paths}

We have emphasized the calculation of probabilities $Pr(PCS_t =i)$ in section 2.1, as the probability distributions of $PCS_t$ for each $t$ give a large-scale description of the Number Pieces. However it should be noted that this description is reductive as it does not consider the possible dynamics between pitch-class sets. The calculation of conditional probabilities $Pr(PCS_{t+\tau} = j | PCS_{t} = i)$ allows a partial description of such dynamics. Yet, considering only these conditional probabilities would amount to assume that the stochastic process is a Markov one, which, as will be made clear below, is false.

To account for the possible relations between pitch-class sets over time, we can study the possible \textit{paths} taken during the performance of a Number Piece. We define a \textit{path} as the set of successive and different pitch-classes sets which can occur during a realization of a Number Piece. For example \{0-1, 1-1, 2-3, 3-7, 2-5, 1-1, 0-1\} is a valid path, while \{0-1, 1-1, 2-3, 2-3, 3-7, 2-5, 1-1, 0-1\} is not as pitch-class set 2-3 is repeated twice. While we can study paths over the whole time of a Number Piece, we will mainly focus on paths on a single time-bracket.

The computer program written for the analysis can be used for the determination of paths, and by averaging over all realizations we can determine their statistic. 

\section{Analysis of \textit{Five}}

\subsection{Structure of the score}

The Number Piece \textit{Five} was written by Cage during 1988 and is dedicated to Wilfried Brennecke and the Wittener Tage \cite{five}. This piece is written for five voices or instruments or mixture of voices and instruments. Each part contains five time-brackets, the third one being fixed. The time-brackets are identical for all players. The temporal structure of \textit{Five} is given on Table \ref{tab:fivetimestruct}.

\begin{table}
{
\begin{center}
\begin{tabular}[c]{ | c | c | c | }

\hline
Time-Bracket & Starting time interval & Ending time interval \\ [1ex] 
\hline              

1 & $[0,450]$ & $[300,750]$ \\
2 & $[600,1050]$ & $[900,1350]$ \\
3 & 1350 & 1650 \\
4 & $[1650,2100]$ & $[1950,2400]$ \\
5 & $[2250,2700]$ & $[2550,3000]$ \\

\hline
\end{tabular}
\end{center}
}
\caption{Temporal structure of the time-brackets of \textit{Five}}
\label{tab:fivetimestruct}
\end{table}

The pitch class structure of each time-bracket is also given in Table \ref{tab:fivepitchstruct}. Each colored circle in the usual circle of semitones represent one player's pitch content for the considered time-bracket, according to the caption given on top. When a player's time-bracket contains multiple pitches, their order is given by the associated numbers, and the pauses and slurs indications are given by the diagrams on the right.

\begin{table}
{
\begin{center}
\begin{tabular}[t]{ | c | c |  }
\hline
Time-Bracket & \shortstack{ Pitch-class content \\ \\ \includegraphics[scale=0.4]{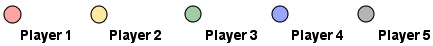} } \\ [1ex] 
\hline   
& \\
1 & \includegraphics[scale=0.4]{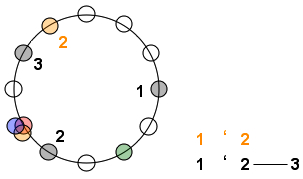} \\
& \\
2 & \includegraphics[scale=0.4]{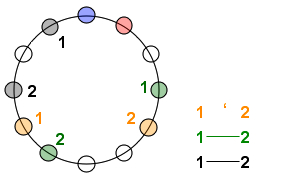} \\
& \\
3 & \includegraphics[scale=0.4]{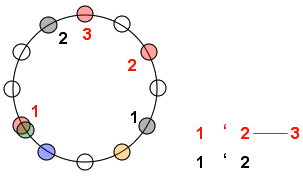} \\
& \\
4 & \includegraphics[scale=0.4]{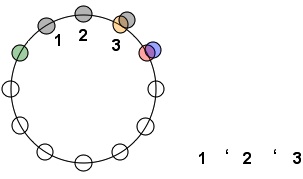} \\
& \\
5 & \includegraphics[scale=0.4]{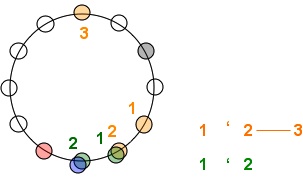} \\

\hline
\end{tabular}
\end{center}
}
\caption{Pitch-class structure of \textit{Five}, represented on the usual circle of semitones. When a time-bracket in a part contains multiple pitches (ordered by the represented numbers), the diagram on the right indicates whether they are separated by pauses (') or slurs (-).}
\label{tab:fivepitchstruct}
\end{table}

\subsection{Statistical analysis}

\begin{figure}
\includegraphics[scale=0.53]{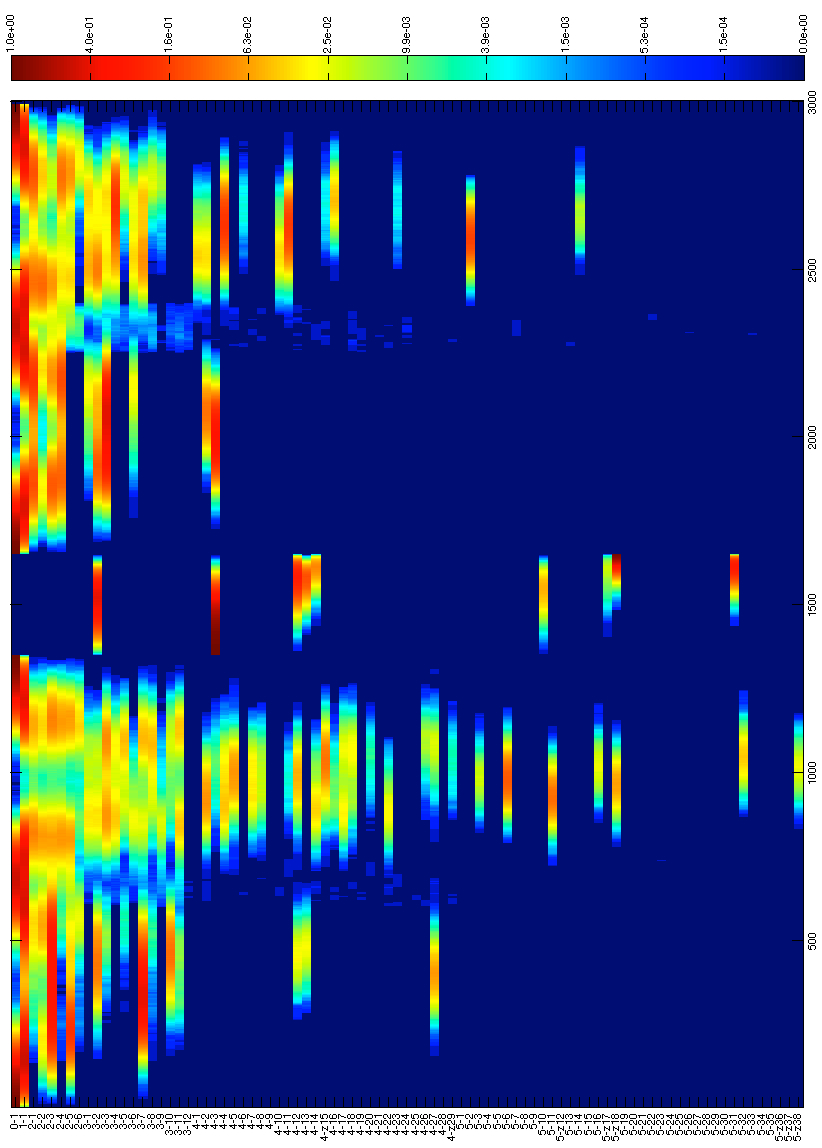}
\caption{Heatmap of the probabilities $Pr(PCS_t =i)$ over the 87 possible pitch-class sets (in ordinate) at each time $t$ (in abscissa) in \textit{Five}. The colorbar indicates the corresponding probabilities in pseudo-logarithmic scale (see text).}
\label{fig:heatmapfive}
\end{figure}

The plot of the probabilities $Pr(PCS_t =i)$ calculated over the 87 possible pitch-class sets (in ordinate) at each time $t$ (in abscissa) is presented in a heatmap plot on Figure \ref{fig:heatmapfive}. Notice that the colorbar, which indicates the corresponding probability values, is given in pseudo-logarithmic scale: if $p$ is a calculated probability and $N$ is the total number of realizations, the color corresponds to the value $1-\dfrac{log_{10}(p+1/N)}{log_{10}(1/N)}$.

As a first observation, one can note that the five different time-brackets are clearly identifiable on this plot.
They are generally separated by either silence or single sounds, as one can see that the probability of obtaining the corresponding pitch-class sets is superior to 0.5 in the external overlaps between the time-brackets. The external overlaps are also characterized by the fact that they contain very rare events. For example, the pitch-class set 4-10 can occur in the external overlap between time-brackets 1 and 2 with a probability of roughly $10^{-4}$. Such rare events correspond to extreme results in the random selection procedure, such as pitches changing at the very end of a time-bracket and so on.

The fixed time-bracket is also clearly visible in this plot, and is characterized by the absence of silence, single sounds or dyads. Indeed, since all players are supposed to start at the same time, there cannot be any such event inside the time-bracket.

It can also be noted that the time-brackets are very different with respect to their possible content. The first time-bracket is characterized by the appearance of only seven possible triads, and just three possible tetrachords. Indeed, it is easy to check from Table \ref{tab:fivepitchstruct} that no pentachord can occur even if the players are all playing at the same time. Time-bracket 2, \textit{a contrario}, is characterized by the appearance of all possible triads except 3-12, many tetrachords with similar probability values, and seven possible pentachords. As said before, time-bracket 3 is characterized by only one possible triad, four tetrachords and four pentachords. The material of time-bracket 4 is reduced, with only four possible triads and two tetrachords. as can be verified on Table \ref{tab:fivepitchstruct}. Finally time-bracket 5 is of moderate complexity between time-bracket 1 and 2.

Another feature shown in this heatmap is that, while many $n$-chords are possible in each time bracket, they do not all occur with the same probabilities, nor at the same time. For example, pitch-class sets 2-3 and 3-7 are prevalent throughout time-bracket 1. Pitch-class set 2-5 may occur at the beginning of this time-bracket with a high probability, while pitch-class sets 3-2 and 3-10 are more probable towards its end. In time-bracket 2, the distributions of $PCS_t$ seem more uniformly spread over the possible pitch-class sets. Yet we can see that pitch-class sets 4-z15, 5-6 and 5-11 have higher probabilities of occurence. A the same time, some pitch-class sets may appear at the beginning or end but are less probable in the middle of the time-bracket, such as 3-3. The peculiar time-structure of time-bracket 3 imposes to begin with pitch-class set 4-3 and to end with pitch-class set 5-z18. Time-bracket 4 is dominated by the high probabilities of obtaining pitch-class sets 3-3 and 4-3. Finally, time-bracket 5 shows higher probabilities of obtaining pitch-class sets 5-2, 4-11, 4-4 or 3-4 in the middle of its structure.

In order to get a better insight about the probabilities of occurence of the different pitch-class sets, we will now focus on a specific analysis of the first time-bracket in \textit{Five}. Table \ref{tab:fivemainpcsets} lists the possible pitch-class sets of cardinality superior to 2 which can occur in this time-bracket. Pitch-class set 3-7 (respectively 3-2) may occur in two different ways: they will be designated by 3-7$_\alpha$, 3-7$_\beta$ (resp. 3-2$_\alpha$, 3-2$_\beta$) as indicated in the Table.

\begin{table}
{
\begin{center}
\begin{tabular}[t]{ | c | c | c |  }
\hline
& &  \\

\includegraphics[scale=0.35]{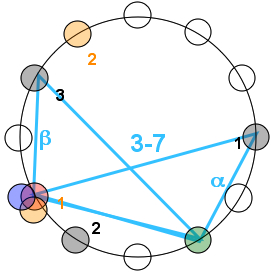} & \includegraphics[scale= 0.35]{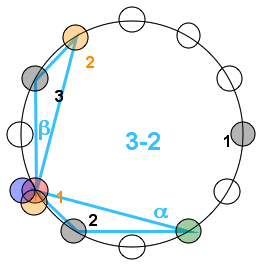} & \includegraphics[scale= 0.35]{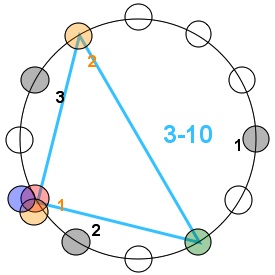} \\

& &  \\
\hline
& &  \\

\includegraphics[scale= 0.35]{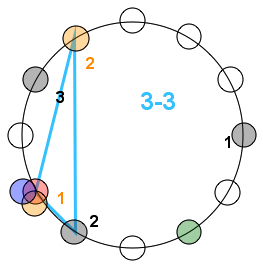} & \includegraphics[scale= 0.35]{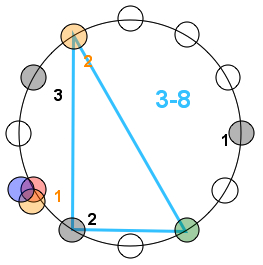} &
\includegraphics[scale= 0.35]{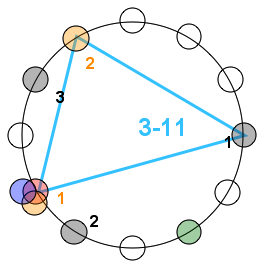} \\

& &  \\
\hline
& &  \\

\includegraphics[scale= 0.35]{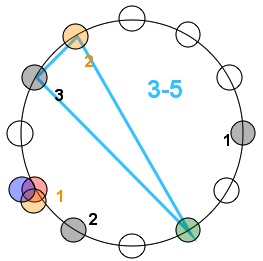} & \includegraphics[scale= 0.35]{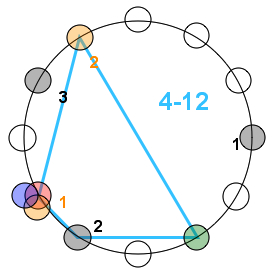} & \includegraphics[scale= 0.35]{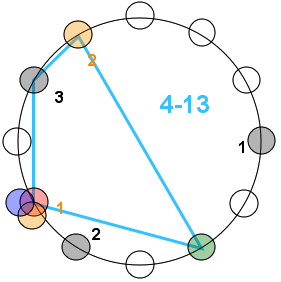} \\

& &  \\
\hline
& &  \\

& \includegraphics[scale= 0.35]{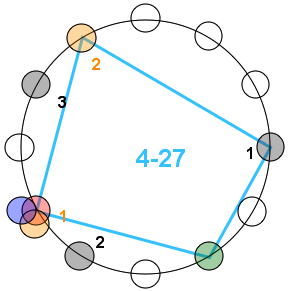} & \\

& &  \\
\hline

\end{tabular}
\end{center}
}
\caption{The possible triads and tetrachords which can occur in the first time-bracket of \textit{Five}. Two possibilities exist in the case of 3-7 and 3-2, which are notated with the indicated symbols $\alpha$ and $\beta$.}
\label{tab:fivemainpcsets}
\end{table}

Most of these chords contain the pitch-class G\#, which can be played by player 1, 2 or 4. From Figure \ref{fig:heatmapfive}, we can see that the chords which contain this pitch-class are more likely to occur in the first time-bracket than the others (this is also valid for dyads). This is rather straightforward as obtaining such a chord requires only one performer playing G\#: the probability of the chord occurence contains the added probabilities of having either player 1, 2 or 4 playing this pitch, hence the increased value. Pitch-class sets 3-5 and 3-8 \textit{in contrario} have very low probabilities of occurence.

Considering the probabilities of the pitch-class sets as represented in Figure \ref{fig:heatmapfive} amounts to studying a static description of the pitch-class set content of time-bracket 1. However, the low probability associated with pitch-class set 3-5 can also be explained from a dynamic point of view by studying the stability of this triad. Indeed, we can see that 3-5 is composed of the second pitch of player 2, the unique pitch of player 3 and the third pitch of player 5. Players 1 and 4 are silent which means either that they have not started their time-bracket yet, or that they have already finished playing it.
The fact that time-brackets coming from separate parts are treated independently in our model implies that the mean of the difference between the starting times (or ending times) of two different players is null. In other terms, the time-brackets start and finish on average at the same time. We thus have two possible evolutions depending on players' 1 and 4 behavior:

\begin{itemize}
\item{Players 1 and 4 have not started their time-bracket. Their entry is therefore imminent since player 2 is already playing its second pitch. The pitch-class set 3-5 is thus unstable and will evolve quickly towards pitch-class set 4-13.}
\item{Players 1 and 4 have finished their time-bracket. All the remaining players are therefore expected to finish soon and the pitch-class set 3-5 is also unstable. It can evolve towards either 2-1, 2-5 or 2-6.}
\end{itemize}

In both cases, we see that the pitch-class set has a short life time, which contributes to its low probability of occurence in time-bracket 1.

\begin{figure}
a. \\
\includegraphics[scale=0.31]{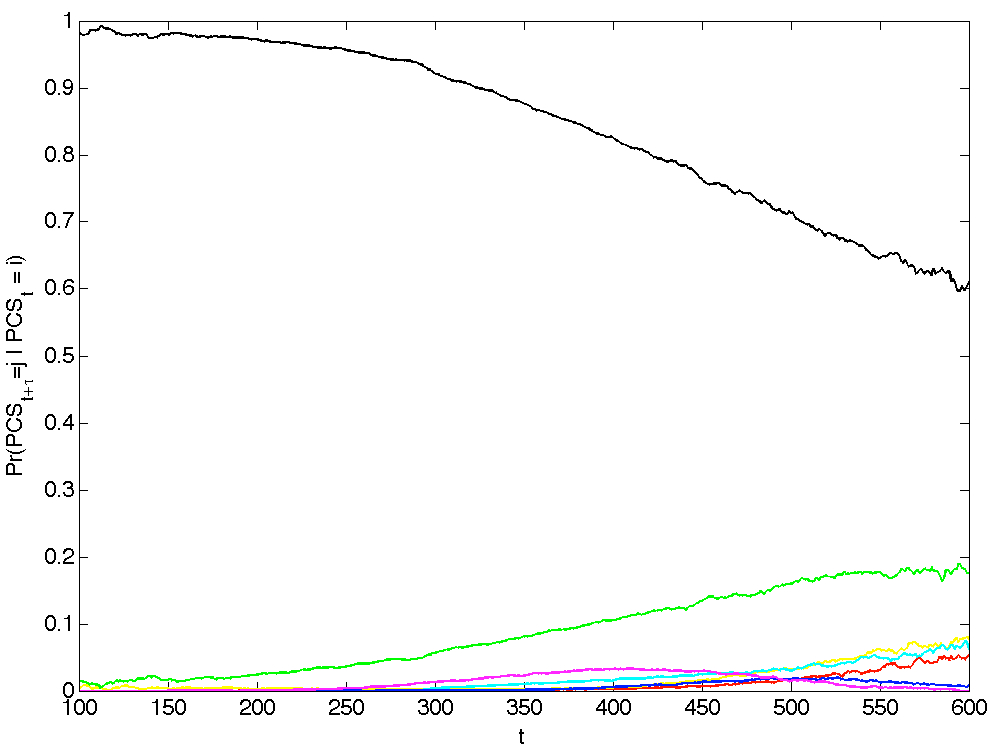} \\
b. \\
\includegraphics[scale=0.31]{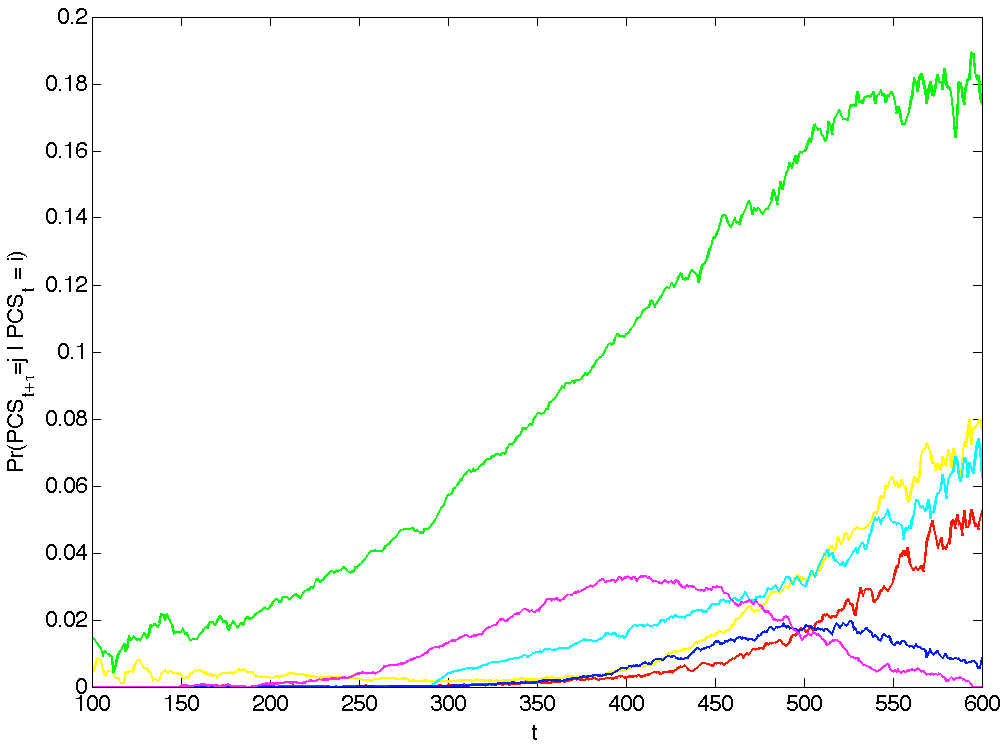}
\caption{Graphs of the probabilities $Pr(PCS_{t+10} = j | PCS_{t} = \text{3-7})$ for different pitch-class sets $j$ over the interval [100,600], calculated over 10$^5$ realizations. Only the pitch-class sets with highest probabilities have been represented: 3-7 (black), 2-3 (green), 4-27 (pink), 2-5 (cyan), 2-2 (yellow), 4-13 (blue), 1-1 (red).  The bottom figure is identical to the top one, with $Pr(PCS_{t+\tau} = \text{3-7} | PCS_{t} = \text{3-7})$ removed for clarity.}
\label{fig:condProbGraphs}
\end{figure}

\begin{table}
{
\small
\begin{center}
\begin{tabular}[t]{ | c | c | }
\hline
Path & Probability \\
\hline
0-1, 1-1, 2-3, 3-7, 2-3, 3-2, 3-7, 2-3, 3-10, 2-3, 1-1, 0-1 & 0.014720 \\ 
0-1, 1-1, 2-5, 3-7, 2-3, 3-2, 3-7, 2-3, 3-10, 2-3, 1-1, 0-1 & 0.010150 \\ 
0-1, 1-1, 2-3, 3-7, 2-3, 3-2, 3-7, 2-3, 1-1, 2-3, 1-1, 0-1 & 0.007070 \\ 
0-1, 1-1, 2-3, 3-7, 2-3, 3-2, 3-7, 2-3, 3-10, 2-3, 1-1, 0-1 & 0.006140 \\ 
0-1, 1-1, 2-3, 3-7, 2-3, 3-2, 3-7, 2-3, 1-1, 0-1, 1-1, 0-1 & 0.004550 \\ 
0-1, 1-1, 2-3, 3-7, 2-3, 3-2, 3-7, 2-3, 3-10, 2-3, 1-1 & 0.004520 \\ 
0-1, 1-1, 2-5, 3-7, 2-3, 3-2, 3-7, 2-3, 1-1, 2-3, 1-1, 0-1 & 0.004480 \\ 
0-1, 1-1, 2-5, 1-1, 2-3, 3-2, 3-7, 2-3, 3-10, 2-3, 1-1, 0-1 & 0.004290 \\ 
0-1, 1-1, 2-5, 3-7, 2-3, 3-2, 3-7, 2-3, 3-10, 2-3, 1-1, 0-1 & 0.004120 \\ 
0-1, 1-1, 2-3, 3-7, 2-3, 3-2, 3-7, 4-13, 3-10, 2-3, 1-1, 0-1 & 0.003760 \\ 
0-1, 1-1, 2-3, 3-7, 4-27, 3-7, 2-3, 3-2, 3-7, 2-3, 1-1, 0-1 & 0.003690 \\ 
0-1, 1-1, 2-3, 3-7, 2-3, 3-2, 3-7, 2-3, 3-10, 2-6, 1-1, 0-1 & 0.003570 \\ 
0-1, 1-1, 2-3, 3-7, 2-3, 3-2, 3-7, 2-3, 1-1, 2-6, 1-1, 0-1 & 0.003410 \\ 
0-1, 1-1, 2-3, 3-7, 2-3, 3-10, 4-12, 4-13, 3-10, 2-3, 1-1, 0-1 & 0.003410 \\ 
0-1, 1-1, 2-5, 3-7, 2-3, 3-2, 3-7, 4-13, 3-10, 2-3, 1-1, 0-1 & 0.003270 \\ 
0-1, 1-1, 2-2, 3-7, 2-3, 3-2, 3-7, 2-3, 3-10, 2-3, 1-1, 0-1 & 0.003200 \\ 
0-1, 1-1, 2-5, 3-7, 2-3, 3-2, 3-7, 2-3, 3-10, 2-3, 1-1 & 0.003070 \\ 
0-1, 1-1, 2-3, 3-7, 4-27, 3-10, 4-12, 4-13, 3-10, 2-3, 1-1, 0-1 & 0.002940 \\ 
0-1, 1-1, 2-3, 3-7, 2-3, 3-2, 3-7, 2-3, 1-1, 2-3, 1-1, 0-1 & 0.002860 \\ 
0-1, 1-1, 2-5, 3-7, 2-3, 3-2, 3-7, 2-3, 1-1, 0-1, 1-1, 0-1 & 0.002830 \\ 
0-1, 1-1, 2-3, 3-7, 2-3, 3-2, 4-12, 4-13, 3-10, 2-3, 1-1, 0-1 & 0.002590 \\ 
0-1, 1-1, 2-3, 3-7, 4-27, 3-10, 2-3, 3-2, 3-7, 2-3, 1-1, 0-1 & 0.002590 \\ 
0-1, 1-1, 2-5, 3-7, 2-3, 3-2, 3-7, 2-3, 3-10, 2-6, 1-1, 0-1 & 0.002580 \\ 
0-1, 1-1, 2-5, 3-7, 2-3, 3-10, 4-12, 4-13, 3-10, 2-3, 1-1, 0-1 & 0.002470 \\ 
0-1, 1-1, 2-3, 3-7, 2-3, 3-2, 3-7, 2-3, 1-1, 2-3, 1-1 & 0.002330 \\ 
0-1, 1-1, 2-3, 3-7, 2-3, 3-10, 2-3, 3-2, 3-7, 2-3, 1-1, 0-1 & 0.002310 \\ 
0-1, 1-1, 2-5, 1-1, 2-1, 2-2, 1-1, 2-3, 3-10, 2-3, 1-1, 0-1 & 0.002200 \\ 
0-1, 1-1, 2-5, 3-7, 2-3, 3-2, 3-7, 2-3, 1-1, 2-6, 1-1, 0-1 & 0.002180 \\ 
0-1, 1-1, 2-5, 3-7, 2-3, 3-2, 4-12, 4-13, 3-10, 2-3, 1-1, 0-1 & 0.002160 \\ 
0-1, 1-1, 2-5, 3-7, 2-3, 3-10, 2-3, 3-2, 3-7, 2-3, 1-1, 0-1 & 0.002140 \\ 
0-1, 1-1, 2-3, 3-7, 2-3, 3-2, 3-7, 2-3, 1-1, 0-1 & 0.002120 \\ 
0-1, 1-1, 2-3, 3-7, 2-3, 3-2, 3-7, 2-3, 1-1, 0-1, 1-1, 0-1 & 0.002030 \\ 
0-1, 1-1, 2-5, 3-7, 4-27, 3-7, 2-3, 3-2, 3-7, 2-3, 1-1, 0-1 & 0.001940 \\ 
0-1, 1-1, 2-5, 1-1, 2-3, 3-2, 3-7, 2-3, 1-1, 2-3, 1-1, 0-1 & 0.001860 \\ 
0-1, 1-1, 2-3, 3-7, 4-27, 3-7, 2-5, 1-1, 2-1, 2-2, 1-1, 0-1 & 0.001850 \\ 
0-1, 1-1, 2-3, 3-7, 2-3, 1-1, 2-1, 2-2, 1-1, 2-3, 1-1, 0-1 & 0.001830 \\ 
0-1, 1-1, 2-3, 3-7, 2-3, 3-2, 3-7, 2-2, 1-1, 2-3, 1-1, 0-1 & 0.001820 \\ 
0-1, 1-1, 2-3, 3-7, 2-5, 1-1, 2-1, 2-2, 1-1, 2-3, 1-1, 0-1 & 0.001820 \\ 
0-1, 1-1, 2-5, 3-7, 2-3, 3-2, 3-7, 2-3, 1-1, 2-3, 1-1, 0-1 & 0.001810 \\ 
0-1, 1-1, 2-3, 3-7, 2-3, 3-2, 3-7, 2-3, 3-10, 2-6, 1-1, 0-1 & 0.001810 \\ 
0-1, 1-1, 2-5, 1-1, 2-3, 3-2, 3-7, 2-3, 3-10, 2-3, 1-1, 0-1 & 0.001800 \\ 
0-1, 1-1, 2-5, 1-1, 2-1, 3-2, 3-7, 2-3, 3-10, 2-3, 1-1, 0-1 & 0.001730 \\ 
0-1, 1-1, 2-3, 3-7, 2-3, 3-2, 3-7, 4-13, 3-10, 2-3, 1-1, 0-1 & 0.001670 \\ 
0-1, 1-1, 2-5, 3-7, 2-3, 3-2, 3-7, 2-2, 1-1, 2-3, 1-1, 0-1 & 0.001640 \\ 
0-1, 1-1, 2-3, 3-7, 4-27, 3-7, 2-3, 3-2, 3-7, 2-3, 1-1, 0-1 & 0.001620 \\ 
0-1, 1-1, 2-3, 3-7, 4-27, 3-10, 2-3, 1-1, 2-1, 2-2, 1-1, 0-1 & 0.001610 \\ 
0-1, 1-1, 2-2, 3-7, 2-3, 3-2, 3-7, 2-3, 1-1, 2-3, 1-1, 0-1 & 0.001590 \\ 
0-1, 1-1, 2-5, 3-7, 4-27, 3-10, 2-3, 3-2, 3-7, 2-3, 1-1, 0-1 & 0.001590 \\ 
0-1, 1-1, 2-3, 3-7, 4-27, 3-7, 2-3, 1-1, 2-1, 2-2, 1-1, 0-1 & 0.001550 \\ 
0-1, 1-1, 2-5, 3-7, 2-3, 1-1, 2-1, 2-2, 1-1, 2-3, 1-1, 0-1 & 0.001550 \\ 
\hline

\end{tabular}
\end{center}
}
\caption{List of the fifty paths with highest probabilities (right) in time-bracket 1.}
\label{tab:fivepaths}
\end{table}

The evolution of chords can be given a more quantitative treatment by studying the conditional probabilities $Pr(PCS_{t+\tau} = j | PCS_{t} = i)$. As stated in Section 2, these probabilities $Pr(PCS_{t+\tau} = j | PCS_{t} = i)$ (with $\tau > 0$) express the probability of transitioning from pitch-class set $i$ to pitch-class $j$ at $\tau$ times in the future. Since this is not a Markov process, these probabilities depend on the time $t$. Consider for example the evolution of pitch-class set 3-7, which is prevalent throughout the time-bracket. Figure \ref{fig:condProbGraphs} presents the graphs of $Pr(PCS_{t+10} = j | PCS_{t} = \text{3-7})$ over the time interval [100,600] for the pitch-class sets 3-7, 2-5, 4-27, 2-7, 2-2, 4-13 and 1-1. The probability $Pr(PCS_{t+10} = \text{3-7} | PCS_{t} = \text{3-7})$ is the highest, meaning that there is a high chance of finding the same chord at one second in the future. However this probability decreases over time as the chord become more and more unstable. The bottom part of Figure \ref{fig:condProbGraphs} shows the probability of transitioning to other chords. The second highest probability corresponds to the transition from 3-7 to 2-3. Indeed, if we are listening to pitch-class set 3-7$_\alpha$, there is a high probability that player 5 would transition from his first pitch to its second, ultimately ending with pitch-class set 3-2. Since there is a pause between these two pitches, we would hear pitch-class set 2-3 played on pitch-classes F and G\#. In the middle of the time-bracket, the transition from 3-7 to 4-27 is also possible, if player 2 starts playing its second pitch. However, this is associated with a lower probability since this requires that player 1 maintains the playing of pitch-class D\#. Towards the end of the time-bracket, some transitions become more and more probable, such as 3-7 to 2-2, 2-5, 2-7 or 1-1. Since we are more likely to encounter pitch-class set 3-7$_\beta$ at this time, these transitions correspond to players who have finished playing their time-bracket.

Incidentally, the above discussion makes it clear that the stochastic process at hand should not be assimilated to a zero- or first-order Markov process. In the case of pitch-class set 3-7, we have seen that its evolution is determined by all the past events and depends on the time considered. To account for the entire past of a pitch-class set, we can study the possible paths inside a time-bracket. Table \ref{tab:fivepaths} lists the fifty paths with highest probabilities occuring in time-bracket 1, calculated over 10$^5$ realizations (27225 different paths were found in total). It can be seen that some of the paths are merely variants of others, and that they confirm the prevalence of pitch-class sets 3-7, 3-2, 3-10 and 4-27 inside this time-bracket.

\begin{figure}
\centering
\includegraphics[scale=0.35]{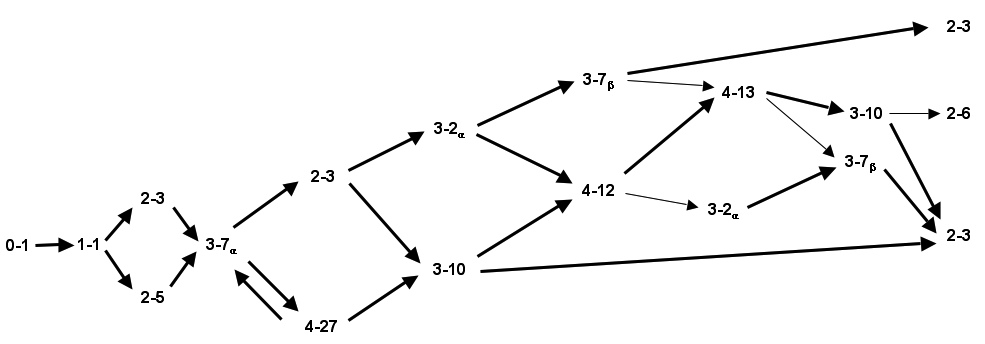} \\

\caption{Model for pitch-class set evolution in the first time-bracket of \textit{Five}. This model represents only the most probable transitions between pitch-class sets. Lines of reduced thickness indicate less probable transitions. Single sounds and silence have been omitted at the end of the model. Pitch-class set 4-27 may evolve to 3-7, in which case the only evolution possible is 2-3.}
\label{fig:fivetb1network}
\end{figure}

With the knowledge of the possible paths and the transition probabilities $Pr(PCS_{t+\tau} = j | PCS_{t} = i)$, we can propose a simplified model for pitch-class set evolution in time-bracket 1. This model, which is shown on Figure \ref{fig:fivetb1network}, considers only the most probable transitions between pitch-class sets. We would like to highlight the fact that, as represented, the model suggests there is a reversible transition between pitch-class sets 4-27 and 3-7. As we have seen above, this assertion is false and while the evolution from 3-7 to 4-27 is reversible, the evolution from 4-27 to 3-7 is not. Though the model is not perfect, it reproduces fairly well the paths in Table \ref{tab:fivepaths}. This network of possible transitions also highlights the recurring role of pitch-class sets 3-7 and 2-3, and to a lesser extent of pitch-class sets 3-2 and 3-10. It also highlights the peculiar role of pitch-class set 4-27: while most of the pitch-classes are accessible through any path in the network, 4-27 is only accessible from 3-7$_\alpha$ towards the beginning of the time-bracket. If 3-7$_\alpha$ transitions instead to 2-3, 3-10 or 3-2$_\alpha$, there is no possibility of hearing 4-27 in the time-bracket anymore. The first time-bracket therefore offers two alternate paths, contributing to the rich dynamics of possibilities during performance.

The same analysis can be carried out for the other time-brackets, though it is likely that the complexity of time-bracket 2, for example, would make it difficult to identify the most probable paths.

\section{Analysis of \textit{Four}}

\subsection{Structure of \textit{Four}}

The Number Piece \textit{Four} was composed by Cage during the same year as \textit{Five}. This Number Piece is written for a string quartet, and is dedicated to the Arditti Quartet \cite{four}.
The piece consists in three sections A, B and C of five minutes each. Each section contains a part for each member, and each part contains ten time-brackets, with the exception of one part in section A in which two time-brackets have been merged into one. Among the ten time-brackets, one of them is always fixed. The parts in each section are not associated with an instrument in particular, as they can be played by any of the players. A performance of \textit{Four} should last either 10, 20 or 30 minutes. If the performance lasts 10 minutes, all players play section B, exchange their parts, then play B again. If a performance duration of 20 minutes has been chosen, players will play sections A and C without pause, exchange their parts, then play sections A and C again. In the last case, players should play sections A, then B and C.

\subsection{Analysis}

Using the same methodology as for \textit{Five} we study the distribution of possible pitch-class sets in each section, the sections being treated independently from one another. Since no individual time-bracket contain chords in any of the sections, the cardinality of the possible pitch-class sets is therefore limited to 4, which encompasses 49 different pitch-class sets from silence to tetrachords.

\begin{figure}
\includegraphics[scale=0.50]{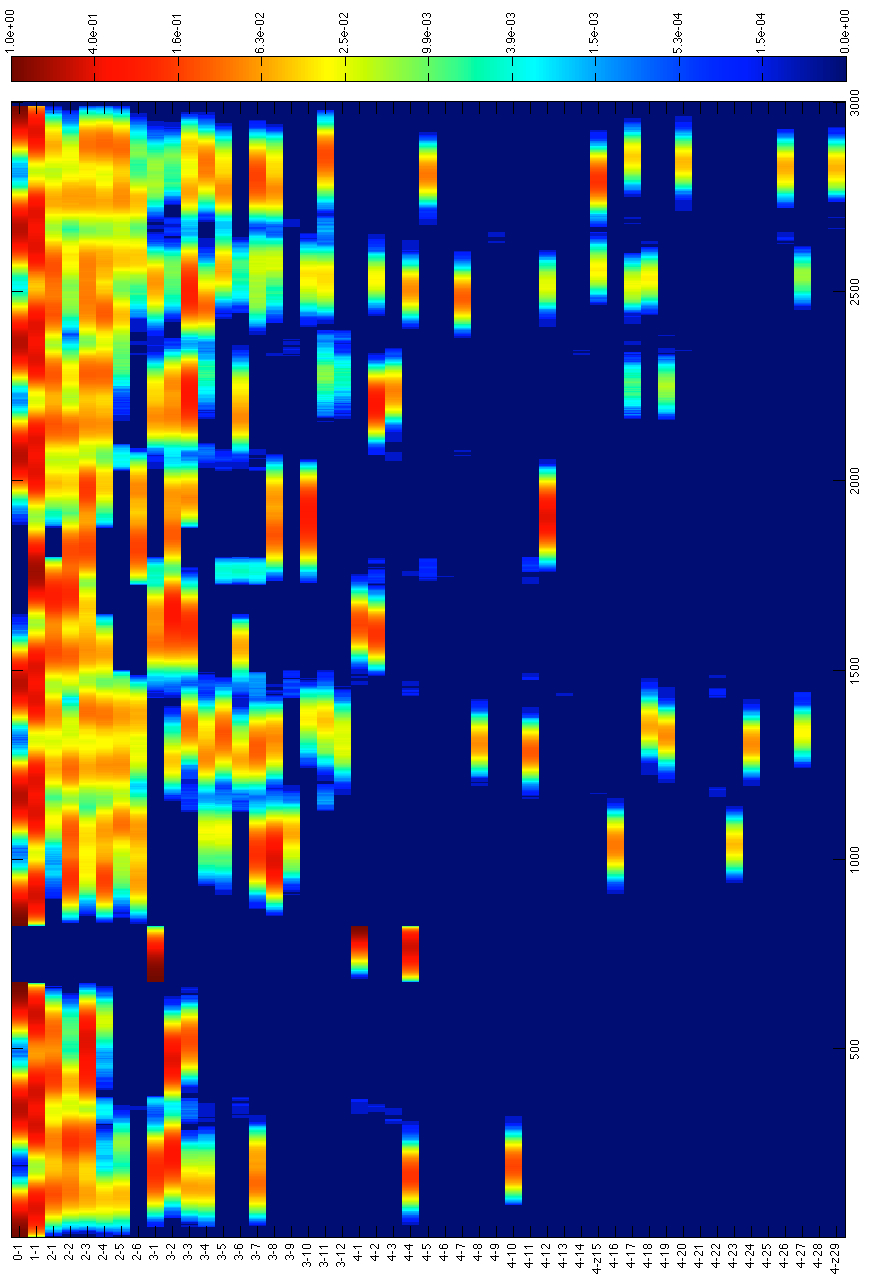}
\caption{Heatmap of the probabilities $Pr(PCS_t =i)$ over the 49 possible pitch-class sets (in ordinate) at each time $t$ (in abscissa) in \textit{Four}, section A. The colorbar indicates the corresponding probabilities in pseudo-logarithmic scale (see text).}
\label{fig:heatmapfourPA}
\end{figure}

\begin{figure}
\includegraphics[scale=0.50]{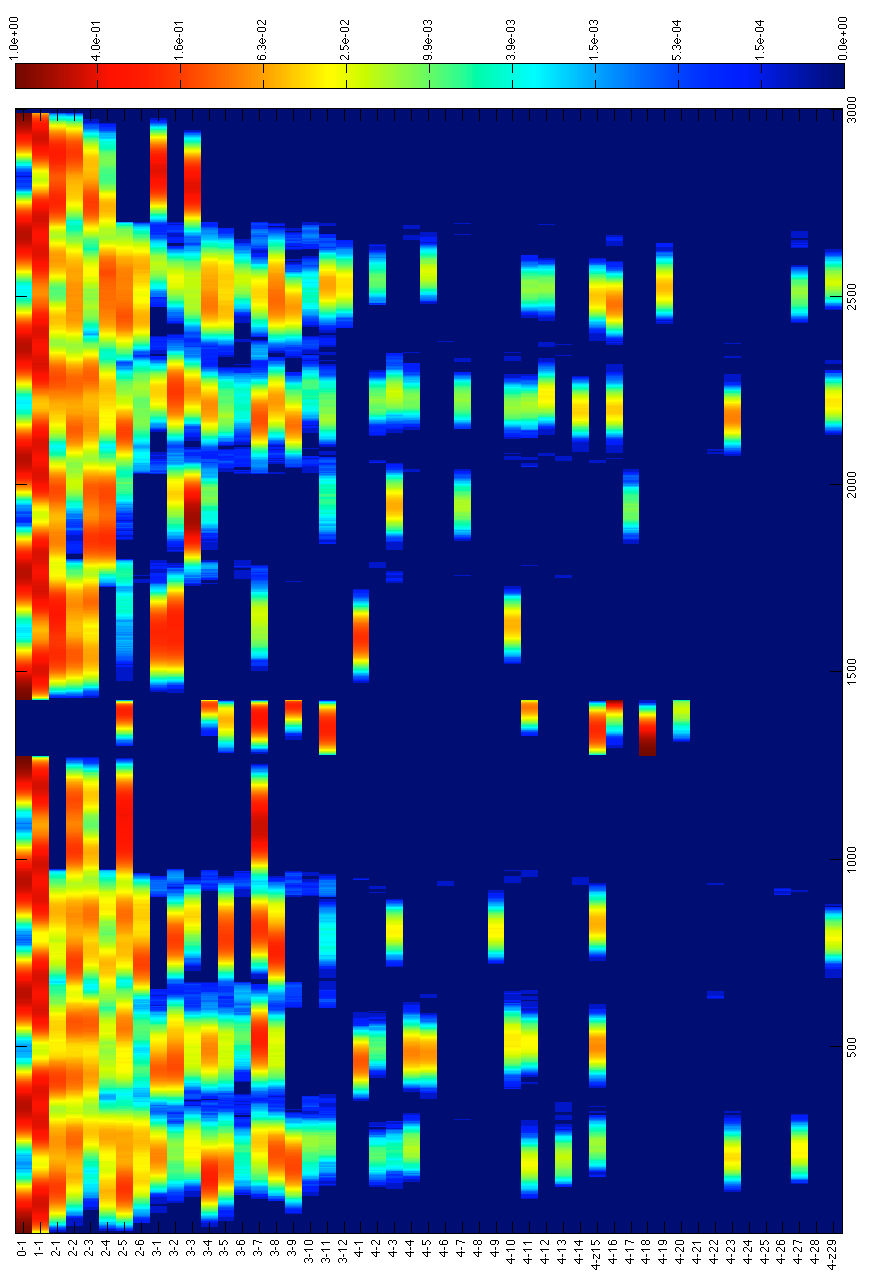}
\caption{Heatmap of the probabilities $Pr(PCS_t =i)$ over the 49 possible pitch-class sets (in ordinate) at each time $t$ (in abscissa) in \textit{Four}, section B. The colorbar indicates the corresponding probabilities in pseudo-logarithmic scale (see text).}
\label{fig:heatmapfourPB}
\end{figure}

\begin{figure}
\includegraphics[scale=0.50]{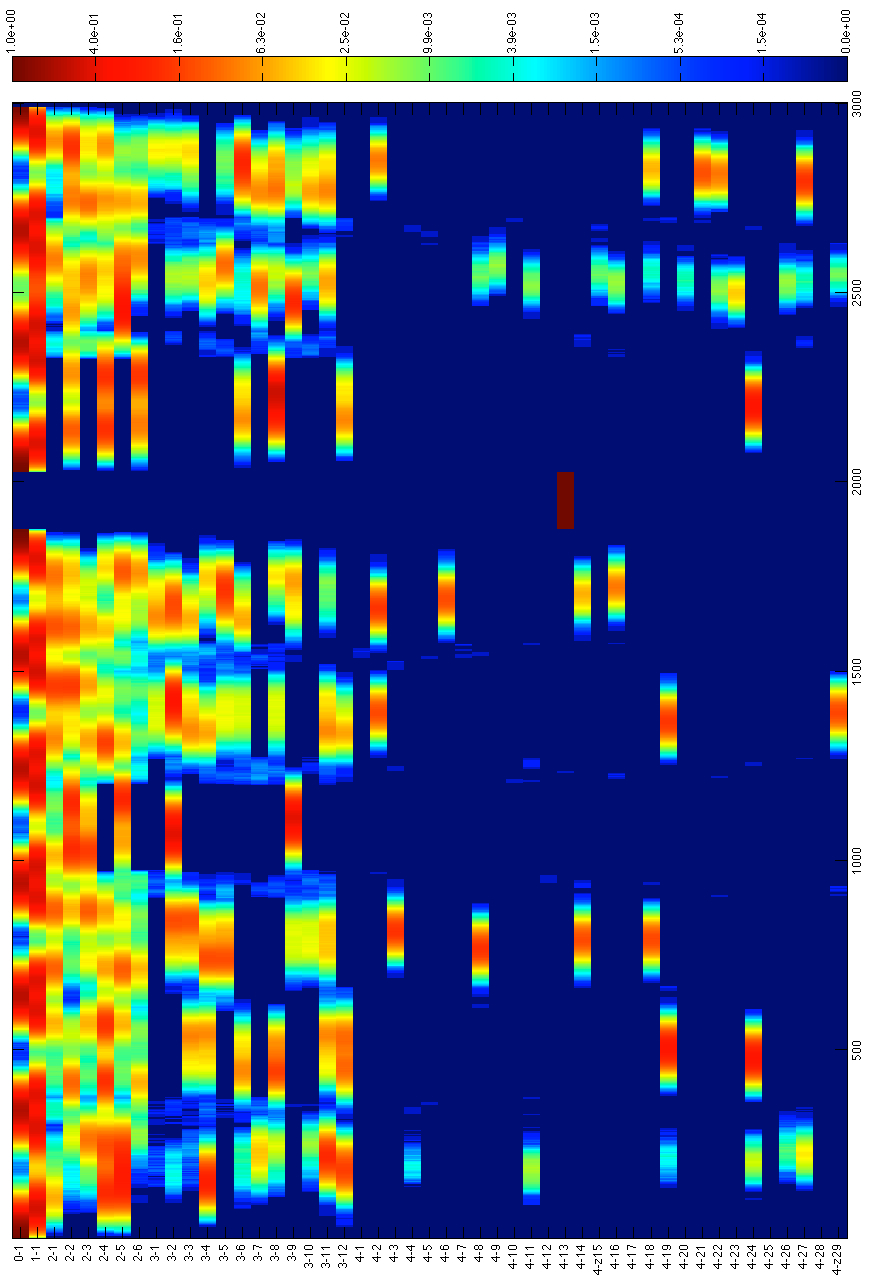}
\caption{Heatmap of the probabilities $Pr(PCS_t =i)$ over the 49 possible pitch-class sets (in ordinate) at each time $t$ (in abscissa) in \textit{Four}, section C. The colorbar indicates the corresponding probabilities in pseudo-logarithmic scale (see text).}
\label{fig:heatmapfourPC}
\end{figure}

The heatmaps of the probabilities $Pr(PCS_t =i)$ calculated over the 49 possible pitch-class sets (in ordinate) at each time $t$ (in abscissa) is presented for each section on Figures \ref{fig:heatmapfourPA}, \ref{fig:heatmapfourPB} and \ref{fig:heatmapfourPC}. As before, the colorbar is given in pseudo-logarithmic scale. Similarly to \textit{Five}, we can readily identify the location of each time-bracket in these heatmaps, as their sonic content is well separated from each other.

Weisser has emphasized about the level consonance heard in \textit{Four}, focusing on major/minor triads (pitch-class set 3-11) and seventh chords, among which the dominant seventh (4-27, which also corresponds to the half-diminished seventh), the major seventh (4-20), and the minor seventh (4-26). The plots presented here allow to quantify this level by looking at the corresponding probabilities of occurence.

Section A is characterized by the scarcity of major/minor triads, as pitch-class set 3-11 is absent from six time-brackets, barely present in one, rare in two, and prevalent in the last time-bracket. Incidentally, this last time-bracket also exhibits moderate probabilities for pitch-class sets 4-20 and 4-26, though there is a higher probability of obtaining the all-interval tetrachord 4-z15. The tetrachords 4-20 and 4-26 are absent from all other time-brackets, and dominant sevenths occur only rarely in time-brackets 5 and 9.

The situation is identical for section B, in which major/minor triads are only prevalent in the fixed time-bracket, more rarely heard in time-bracket 9, and barely present in the remaining four time-brackets where they occur. It is interesting to note that the possible minor triad which can occur in the fixed time-bracket as been categorized as part of a "thwarted triad/seventh chord" by Weisser (see Example 10.c in \cite{weisser2}), missing the possibility that this minor triad could exist on its own if player 4 has started playing its second pitch before any other player. Pitch-class set 4-26 is absent from the entire section, while 4-27 has only two low-probability occurences, and 4-20 only one.

Section C stands again the other sections, given the greater possibilities of hearing pitch-class set 3-11, with occurences in seven out of the ten time-brackets. However, pitch-class set 4-20 is virtually absent from the section, similarly to pitch-class set 4-26. This is almost the same for pitch-class set 4-27, except for the last time-bracket where it is prevalent.

\section{Conclusions}

We have shown in this paper that a statistical approach of the possible pitch-class set in the Number Pieces of John Cage, drawn from previous studies on single time-brackets, allows to analyze at once the distribution of sonic content during a performance and to quantify the relative probabilities of occurence of the chords in each time-bracket.
The analyses proposed in this paper rely on the hypotheses which were exposed in the methodology section regarding the procedures used for selecting time marks for the time-brackets. A number of issues can therefore raised concerning this particular model:

\begin{itemize}
\item{The outer limits of the time-brackets are chosen through a random selection within the given intervals. We have discussed in \cite{popoff2} the possible shortcomings of this approach, among which the fact that it may not represent human behavior accurately, since humans are poor random generators and are influenced by the surrounding stimuli as well as previous events.}
\item{The selection of multiple pitches inside a time-bracket is made through a succession of random time-mark choices. Again, this might not be representative of human behavior. We have argued in \cite{popoff1} that two conceptions of time, based either on real time measurements or on time differences (durations), compete in the Number Pieces, and a selection based on durations may be more appropriate.}
\item{Parts belonging to different players are treated independently. However, musicians are very likely to listen to each other during a performance of a Number Piece and to take different decisions depending on what they perceive. }
\end{itemize}

The modelisation of the Number Pieces is therefore an open problem, which would benefit from a more thorough investigation of actual musician's behavior. In particular, it would be interesting to see if modifications in the above points (for example taking into account cooperative playing) would influence the results presented above and to what extent. Note however that the approach used here has the advantage of simplicity for computer implementation, and has also been used for the automated computer generation of performances of the Number Pieces (\cite{sluchin}).

Finally, we wish to underline that we have considered here the probability distributions of the possible pitch-class sets without discussing their perceived consonant, dissonant or even tonal nature. Parncutt (\cite{parncutt}) has emphasized the fact that pitch-class sets may have tonal implications, drawing from the work by Krumhansl (\cite{krum1}) on the perceived nature of musical pitch. Since we have access to the instantaneous pitch-class content of the Number Pieces and by using Krumhansl key-finding algorithm (\cite{krum1}, \cite{krum2}), it would be interesting to track the possible keys evoked throughout a performance of a Number Piece, and even to study the distribution of these keys over a large number of realizations.


\label{lastpage}


\begin{thebibliography}{}

\bibitem[1]{weisser1}
B. J. Weisser, "Notational Practice in Contemporary Music: A Critique of Three Compositional Models (Luciano Berio, John Cage and Brian Ferneyrough)" (Ph.D. dissertation, City University of New York, 1998), pp. 82-83.

\bibitem[2]{weisser2}
B. Weisser, "John Cage: '... The Whole Paper Would Potentially Be Sound': Time-Brackets and The Number Pieces (1981-92)", Perspectives of New Music, 41(2), pp. 176-225.

\bibitem[3]{haskins}
R. Haskins, "An Anarchic Society of Sounds: The Number Pieces of John Cage" (Ph.D. dissertation, University of Rochester, New York, 2004), pp. 245.

\bibitem[4]{cage}
J. Cage and J. Retallack, "Musicage: Cage Muses on Words, Art, Music. John Cage in conversation with Joan Retallack", ed. Joan Retallack, Hanover, NH: University Press of New England, Wesleyan University Press, 1996, p. 108

\bibitem[5]{popoff1}
A. Popoff, "John CageÕ's Number Pieces: The Meta-Structure of Time-Brackets and the Notion of Time", Perspectives of New Music, 48 (1), pp. 65Ð84

\bibitem[6]{popoff2}
A. Popoff, "Indeterminate Music and Probability Spaces: The Case of John Cage's Number Pieces", Proceedings of the Mathematics and Computation in Music - Third International
               Conference, LNAI 6726, Springer, 2011, pp. 220-229

\bibitem[7]{starr1}
D. Starr, "Sets, Invariance and Partitions", \textit{Journal of Music Theory}, 22 (1), pp. 1Ð42

\bibitem[8]{starr2}
A. R. Brinkman, "Pascal Programming for Music Research", University of Chicago Press, 1990, p. 629

\bibitem[9]{forte}
A. Forte, "The Structure of Atonal Music", New Haven and London: Yale University Press, 1973

\bibitem[10]{five}
John Cage, "Five", New York: C.F. Peters, 1988, EP 67214, performance notes

\bibitem[11]{four}
John Cage, "Four", New York: C.F. Peters, 1988, EP 67304, performance notes

\bibitem[12]{sluchin}
B. Sluchin, M. Malt, "A computer aided interpretation interface for John CageÕs number piece Two$^5$", Actes des JournŽes dÕInformatique Musicale (JIM 2012), Mons, Belgique, 2012

\bibitem[13]{parncutt}
R. Parncutt, "Tonal Implications of Harmonic and Melodic T$_n$-types", {Proceedings of the Mathematics and Computation in Music - First International
               Conference}, CCIS 37, Springer, 2007, pp. 124-137

\bibitem[14]{krum1}
C.L. Krumhansl, "Cognitive Foundations of Musical Pitch", Oxford University Press, New York, 1990

\bibitem[15]{krum2}
P. Toiviainen, C.L. Krumhansl, "Measuring and modeling real-time responses to music: Tonality Induction", Perception, 32, pp. 741-766

\bibitem[16]{krum3}
D. Temperly, "What's Key for Key? The Krumhansl-Schmuckler Key-Finding Algorithm Reconsidered", Music Perception, 17 (1), pp. 65-100

\end{thebibliography}
\end{document}